\documentclass[twocolumn]{emulateapj}

\slugcomment{Submitted, \today}

\shorttitle{Planck-Scale-Induced Blurring}
\shortauthors{Steinbring}

\input epsf
\def\plotone#1{\centering \leavevmode
\epsfxsize=1.0\columnwidth \epsfbox{#1}}
\def\plotonewide#1{\centering \leavevmode
\epsfxsize=1.5\columnwidth \epsfbox{#1}}

\begin{document}

\title{Detectability of Planck-Scale-Induced Blurring with Gamma-Ray Bursts}

\author{Eric Steinbring\altaffilmark{1}}

\altaffiltext{1}{National Research Council Canada, Herzberg Astronomy and Astrophysics, Victoria, BC V9E 2E7, Canada}

\begin{abstract}
Microscopic fluctuations inherent to the fuzziness of spacetime at the Planck scale might accumulate in wavefronts propagating a cosmological distance and lead to noticeable blurring in an image of a pointlike source. Distant quasars viewed in the optical and ultraviolet with {\it Hubble Space Telescope} ({\it HST}) may show this weakly, and if real suggests a stronger effect should be seen for Gamma-Ray Bursts (GRBs) in X-rays and $\gamma$-rays. Those telescopes, however, operate far from their diffraction limits. A description of how Planck-scale-induced blurring could be sensed at high energy, even with cosmic rays, while still agreeing with the {\it HST} results is discussed. It predicts dilated apparent source size and inflated uncertainties in positional centroids, effectively a threshold angular accuracy restricting knowledge of source location on the sky. These outcomes are found to be consistent with an analysis of the 10 highest-redshift GRB detections reported for the {\it Fermi} satellite. Confusion with photon cascade and scattering phenomena is also possible; prospects for a definitive multiwavelength measurement are considered.
\end{abstract}

\keywords{gravitation --- gamma rays: bursts}

\section{Introduction}\label{introduction}
 
Whether astronomical images show tangible evidence of electromagnetic waves having transited through the spacetime foam is an intriguing possibility based on a simple premise: if space is not smooth, travel along a lightpath must be subject to continual, random distance fluctuations $\pm \delta l$ proportional to Planck length $l_{\rm P} \sim {10}^{-35}$ m (or equivalently, timescale $t_{\rm P}\sim 10^{-44}~{\rm s}$) which accumulated over a sufficient distance will erode the phase coherence of wavefronts \citep{Lieu2003}. The expected effect - blurring of a pointlike source beyond that explainable by other means - has not been decisively ruled out in searches with {\it Hubble Space Telescope} ({\it HST}) among distant galaxies \citep{Ng2003, Ragazzoni2003} and quasars \citep{Steinbring2007, Christiansen2011, Perlman2011, Tamburini2011}.

Blurring would constitute a fundamental uncertainty in particle localization. Although general relativity neglects the quantum nature of particles, various forms of quantum gravity, including string theory, demand irreducable fluctuations at the Planck scale of $E_{\rm P}\sim 10^{28}~{\rm eV}$; phase decoherence is just one phenomenology that may allow tests of those \citep[see][for a recent review]{Amelino-Camelia2013}. Among the best constrained is compliance with Lorentz invariance, uniformity of the speed of light with energy. Limits on time-of-flight delays from Gamma-Ray Bursts (GRBs) make use of the long distance over which those photons travel, with energy dispersion $\delta E$ scaling as $l/c$ \citep{Amelino-Camelia1998}. For example, measuments of GRB 090510 ($z=0.9$) by \cite{Abdo2009} are inconsistent with quantum-gravity formulations which fall outside a strict lower limit of $1.2 E_{\rm P}$ (or inversely with wavelength) on any linear energy dependence on the speed of light.

Similarily, the strength of phase degradation at observed wavelength $\lambda$ depends on the summation of phase perturbations $\Delta \phi = 2\pi \delta l/\lambda$ along a given trajectory. Following \cite{Steinbring2007} for the luminosity distance of $L = ({c/{H_0 q_0^2}})[q_0 z - (1 - q_0)(\sqrt{1 + 2 q_0 z} - 1)]$ this reaches
$$\Delta\phi_{\rm max}=2\pi a_0 {l_{\rm P}^{\alpha}\over{\lambda}}\Big{\{}\int_0^z L^{1 - \alpha} {\rm d}z + {{(1-\alpha)c}\over{H_0 q_0}} $$
$$~~~~~~~~~~~~~~~~~~~~~~~\times \int_0^z (1+z) L^{-\alpha} \Big{[}1 - {{1 - q_0}\over{\sqrt{1 + 2 q_0 z}}}\Big{]} {\rm d}z \Big{\}}$$
$$~~~~=\Delta\phi_{\rm los} + \Delta \phi_z=(1+z)\Delta\phi_0, \eqno(1)$$
where $q_0={{\Omega_0}/{2}} - {{\Lambda c^2}/{3 H_0^2}}$ is the deceleration parameter\footnote{A cosmology with $\Omega_\Lambda=0.7$, $\Omega_{\rm M}=0.3$, and $H_0=70~{\rm km}~{\rm s}^{-1}~{\rm Mpc}^{-1}$ is assumed throughout.}. The effect is stronger is bluer light; a correction by a factor of $1+z$ due to cosmological expansion is the same expected for Lorentz-invariance violation \citep{Jacob2008}. Here $\Delta\phi_{\rm los}$ includes waves propagating from any point along the line of sight, $\Delta \phi_z$ are exclusively those redshifted to the observer, and
$$\Delta\phi_0=2\pi a_0 {l_{\rm P}^{\alpha}\over{\lambda}}L^{1 - \alpha} \eqno(2)$$ 
is as in \cite{Ng2003} for $a_0\sim 1$ and $\alpha$ specifying the quantum-gravity model: $1/2$ implies a random walk and $2/3$ is consistent with the holographic principle. Choosing the last scales all down by $(l_{\rm P}/L)^{0.17}$; it vanishes for $\alpha=1$. The maximal degree of blurring would be easy to see if it were, say, 50\% of the diffraction limit $1.22\cdot\lambda/D$ for an ideal telescope of clear aperture diameter $D$, effectively making a disk of diameter equal to the first diffraction minimum. But for the same $D$ and a light travel path reduced to $L_{\rm C}=L/(1+z)$, the co-moving distance as advocated by \cite{Perlman2011}, with $a_0=1$, $\alpha=2/3$ and $z=4$ (for a flat cosmology) $\Delta \phi_0$ would swell images to just 6\% beyond diffraction. At this weak level, for any $a_0$, only a lower limit can be probed for $\alpha$.

Instead, Planck-scale-induced blurring may exist but be weaker than any other aberration, and so remain invisible. If the sensible phase error is just the difference in induced phase angle across the aperture, this might be only \citep{Maziashvili2009}
$$\Delta \phi_D = 2 \pi a_0~{\Big{(}{{l_{\rm P}}\over{\lambda}}\Big{)}}^\alpha~\Big{[}1 + \Big{(}{{D}\over{0.16~L}}\Big{)}^\alpha\Big{]}. \eqno(3)$$ 
For the same $a_0$ and $\alpha$ this would be diminished by more than ${10}^{10}$ from $\Delta \phi_0$, much less than diffraction and possibly untestable. But a hard lower limit is presumably 
$$\Delta \phi_{\rm P} = 2\pi{{l_{\rm P}}\over{\lambda}}, \eqno(4)$$ 
a Planck ``resolution" a further ${10}^{10}$ less than $\Delta \phi_D$ in the optical, and minimally $\sim l_{\rm P}/D \approx 10^{-35}~{\rm radians}$ for $D\sim1$ m. No real telescope could approach this hyperfine sharpness, because to do so any optical imperfections would have to be smaller than $l_{\rm P}$. Note that, without precluding any phase error up to the maximum, the ratio between that and the least effect is always $\Delta \phi_{\rm max} / \Delta \phi_{\rm P}=(1 + z) a_0 (L/l_{\rm P})^{1 - \alpha}$, a ``fixed" amplitude with no dependence on $\lambda$.

\begin{figure}
\plotone{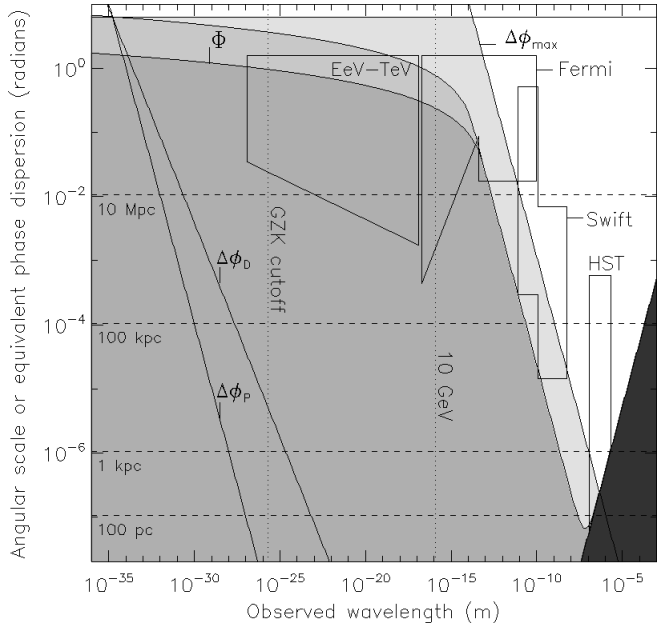}
\caption{Angular source sizes at $z=4$ relative to the diffraction limit of a perfect telescope of aperture $D=2.4$ m (dark shading). Limits imposed by equations 1, 3, and 4 are plotted (other shaded regions are discussed in Section~\ref{description}). Regions visible with {\it HST}, X-ray, $\gamma$-ray satellites and TeV to EeV telescopes are outlined.}\label{figure_blurring_wide}
\end{figure}

The most stringent observational limits on phase decoherence obtained so far come from a sample of 99 Sloan Digital Sky Survey (SDSS) quasars spanning $3.9\leq z\leq 6.3$. These were observed with the {\it HST} Advanced Camera for Surveys High Resolution Channel (ACS/HRC) in filters F775W and F850LP. Blurring in this regime may be well under $0\farcs01$, too weak to be detected directly as a broadening of point-spread function (PSF) full-width at half-maximum (FWHM). But it may be responsible for a remarkable trend in Strehl ratio $S$ - relative peak heights - of diffraction spikes and point sources blurred according to $\Delta \phi_z$ for $a_0=1$ and $\alpha=0.67$. No sources are found to be less blurred than this, nor as much as $\Delta \phi_{\rm max}$. A decrement from $S_0=0.82$ for a synthetic quasar PSF implies a source extended by $\sqrt{S_0/S}\approx \sqrt{0.82/0.72}=1.07$. Using Ultra-Deep Field (UDF) Wide-Field Channel F435W and F606W images of a few (lower-$z$) quasars, \cite{Christiansen2011} report blurring as per $L_{\rm C}$, yielding $\alpha<0.65$. A difficulty with both of these studies is that Active Galactic Nucleii (AGN) have intrinsic diameters on the order of parsecs, and if as large as 100 - 200 pc at $z=4$ would be at a level just approaching that observed, as illustrated in Figure~\ref{figure_blurring_wide}.

A clearer result might be expected from employing GRBs. They are stellar in nature, and although embedded within a host galaxy as are AGN, their image contrast can be greater (temporarily) in the optical/infrared. For a burst lasting 10 s the emission region would be confined to a radius of $10~{\rm s}~\times c \approx 10^{-7}~{\rm pc}$, spanning an immeasurably small $\sim 10^{-11}~{\rm arcsec}$ for a source at $z=0.5$. At X-ray and higher energies this may be superimposed on a diffuse halo $\sim 100~{\rm pc}-1~{\rm kpc}$ across due to dust scattering \citep[][]{Vaughan2004} or possibly $\sim 1 - 10~{\rm Mpc}$ due to the photon cascade from upscattered cosmic-ray background (CMB) photons \citep{Aharonian1994, Takahashi2010}. Sources are routinely located to within a few arcminutes on the sky, e.g. GRB 070125 \citep[$z=1.5$;][]{Bellm2008} obtained indirectly via triangulation of burst signals between X-ray satellites within the Interplanetary Network (IPN). Even so, it would be nearly impossible to distinguish between scattering scenarios and Planck-induced blurring in X-rays as per $\Delta \phi_{\rm max}$ for $\alpha=0.67$, even with a $z=4$ GRB. And $\Delta \phi_0$ would still be invisible for the most distant known \cite[GRB 090423, $z=8.2$][]{Tanvir2009, Salvaterra2009} detected at 0.2 to 150 keV by {\it Swift}.

Looking for blurred GRBs in $\gamma$-rays, or indeed for any compact source at higher energy, then poses a dilemma. If $\alpha=0.67$ and all $\gamma$-rays have the lesser phase dispersion $\Delta \phi_0$, this should be obvious at energies of 60 MeV or more for a GRB with only $z=0.5$, swelling to over 1 radian - for either choice of $L$. This is the regime of the {\it Fermi} observatory Gamma-Ray Burst Monitor (GBM) and Large Area Telescope (LAT). The latter has a resolution of $5^\circ$ at 30 MeV to 1.5\arcmin~at 60 GeV, and has found at least 10 GRBs near 10 GeV with $z\geq0.5$ \citep{Ackermann2013a, Ackermann2013b}. Even if $\Delta \phi_0$ were ${10}^{26}$ times weaker these should be smeared over the entire sky, making identification impossible. Furthermore, unless the blurring mechanism breaks down for cosmic rays, the disagreement with longer wavelength data could be more serious. Only $\Delta \phi_{\rm D}$ and $\Delta \phi_{\rm P}$ can easily avoid conflict with anisotropy near the Greisen-Zatsepin-Kuzmin (GZK) cutoff at $6\times10^{19}$ eV. Beyond this, interaction with the CMB is known to suppress flux \citep{Abbasi2008} and interaction with magnetic fields is important; the cosmic-ray sky is probably not uniform. At least one ``hotspot" with a size between $20^{\circ}$ and $40^{\circ}$ across has recently been reported at EeV energies with the Telescope Array (TA) experiment \citep{Abbasi2014}.

If the optical results are correct, and without excluding any visible blurring at shorter wavelengths, there is a way to reconcile the opposing observations. In Section~\ref{description} the sensed blurring is taken to be an average of all possible phase dispersions relative to a simple model of telescope imaging. A limit of $2\pi$ naturally bounds the effect, avoiding nonphysical phase dispersions at high energy. It will be shown that this prescription agrees with previous {\it HST} AGN results and does not preclude identification of GRBs with {\it Fermi}. In fact, it predicts a degradation in angular resolution near 10 GeV, and a first effort to test that is reported here. The {\it Fermi} data and analysis are presented in Section~\ref{observations}, and are consistent with this new view, although the currently limited sample and potential for partial resolution of sources eludes a definitive detection. Future prospects for observational verification are outlined in Section~\ref{summary}.

\section{Improved Description of Blurring}\label{description}

Previous work avoided searches for blurring at wavelengths shortward of the UV. Details of telescope optics should not be critical; a monochromatic treatment will be sufficient for sources with a well-defined peak flux, such as AGNs or GRBs. Even for a telescope operating far from the diffraction limit the visibility of Planck-scale-induced aberration, and so the ability to disentangle this from an extended source, must still be relative to telescope angular resolution - the finest spatial scales it can differentiate at each wavelength. That will hold true until blurring grows outside the field of view (FoV), where photons are effectively scattered into the background.

\subsection{Distribution of Phase Dispersions}

Those recorded photons need not all correspond to the maximal case of $\Delta \phi_{\rm max}$, nor is any lesser limit fundamental, except $\Delta \phi_{\rm P}$. Moreover, a suitably long exposure plausibly produces an image averaging all detectable phase dispersions, and if those have a distribution with amplitude ${\Delta \phi}~\sigma (\Delta \phi) = 1-A \log({{\Delta \phi}/{\Delta \phi_{\rm P}}})$, this is
$${1\over{A}} \int \Delta \phi ~\sigma (\Delta \phi) ~{\rm d}{\Delta \phi} = (1 + z) \Delta \phi_0, \eqno(6)$$
which recovers equation 1. As mentioned in Section~\ref{introduction}, normalization $A = 1/\log{[(1 + z) a_0 (L/l_{\rm P})^{1 - \alpha}]}$ is constant for all $\lambda$, but inherits a slow dependence on $z$. It is 0.021 at $z=4$ for $a_0=1$ and $\alpha=0.67$. Bound by an upper limit of $2\pi$ the integral (shown as the upper curve in Figure~\ref{figure_blurring_wide}) can converge smoothly to that in the absence of diffraction as $\lambda\rightarrow l_{\rm P}$ and so also matches both $\Delta \phi_{\rm D}$ and $\Delta \phi_{\rm P}$ there. It is worth reiterating that this procedure merely formalizes the linear superposition of waves with phase dispersion $\Delta\phi$. Rescaling by $\Delta\phi_{\rm P}$ is certainly convenient for the summation but does not imply any small-angle scattering approximation. No ``extra" parameter has been added to the theory.

\subsection{Simple Telescope Model}

A rudimentary optical prescription permits direct comparison across the observable spectrum. A useful descriptor is the ratio of smallest resolvable scale to diffraction, $R(\lambda/D)^\rho$ where $\rho$ is the slope of instrument response as a function of wavelength. Thus, in the case of a diffraction-limited telescope: $R=1.22$. The value of $R$ incorporates all instrument aberrations, certainly known to within a factor of 2, realistically to an accuracy to 10\%, but not usually better than 5\%. A FoV of $2\pi$ is admittedly unrealistic, as no single telescope views simultaneously the entire sky, although this may at least be taken as a formal ceiling. It is the limiting case where all incident waves arrive obliquely (just perpendicular) to the telescope beam, even for an opening angle of exactly $\pi$, horizon to horizon. A more practical situation at MeV energies or higher is an angular FoV with a cone of opening angle perhaps $45^\circ$ or as much as $90^\circ$.

\subsection{Integrated Effect}

Consider a point source viewed by a telescope with FoV spanned by angle $\theta$. For $\theta \leq 2\pi$ and $A>0$ the integral in equation 6 implies an observed PSF mean width of
$$\Phi = R\Big{(}{{\lambda}\over{D}}\Big{)}^\rho + \int_0^\theta \Delta \phi ~\sigma (\Delta \phi) ~{\rm d}{\Delta \phi} ~~~~~~~~~~~~~~~~~~~~$$
$$~~~~~~~~~ = \Phi_R + \Phi_\theta = A R \Big{(}{\lambda\over{D}}\Big{)}^\rho \Big{[} 1 + \log{\Big{(}{{2\pi l_{\rm P} D^\rho}\over{R \lambda^{\rho+1}}}\Big{)}}\Big{]}~~~~~~~~~~~~~~~~~~~~~~~~~~~~~~~~~~~~~~~~~~~~~$$
$$~~~~~~~~~~~~~~~~~~~~~~~~~~~~~~~ + \theta \Big{\{} 1 + A\Big{[} 1 + \log{\Big{(}{{2\pi l_{\rm P}}\over{\theta \lambda}}\Big{)}}\Big{]}\Big{\}}, \eqno(7)$$
where the two components arrive from integration by parts, that is, splitting the integral above and below $R(\lambda/D)^\rho$. The $\Phi_{\rm R}$ portion includes all phase dispersions up to the telescope resolution limit. This can be interpreted as indicating the smallest perceivable angular resolution is inflated by on average a factor $1 + A + A\log{(2 \pi l_{\rm P} D^\rho / R \lambda^{\rho+1})}$ up to wavelengths where the maximal blurring is not visible, and can only vanish near the diffraction limit as $\lambda\rightarrow l_{\rm P}$. In other words, there is a range of wavelengths where $\Phi$ must be more than the telescope resolution and less than its FoV.

\subsection{General Comparison Across Observable Spectrum}

This averaging procedure fits the basic picture presented by current observations. It matches the trend for SDSS $3.9 \leq z\leq 6.3$ AGN with {\it HST}, which follows from the realization that $\Delta \phi_0\approx 0.48\Delta\phi_z$ near $z=4$ (for example, by numerical integration of equation 1). For $\Delta \phi_z$ close to the diffraction limit, the PSF must therefore be extended (on average) by a ratio $\Phi/(1.22\lambda/D)=1 + 0.48(1 + z)A$, and so produces the same slope found in \cite{Steinbring2007}; reaching approximately 1.07 at $z=6.3$. And agreement continues to much shorter wavelengths, as illustrated in Figure~\ref{figure_blurring_wide}. Equation 7 is plotted for $a_0=1$, $\alpha=0.67$, $D=2.4$ m, and $z=4$. The lightly shaded region indicates where all photons would reach the limit set by equation 1. The darkest-shaded region is that precluded by the diffraction limit of a perfect $2.4$-m diameter telescope, although $\Phi_\theta$ is only weakly dependent on $D$ for X-rays or shorter wavelengths, and the differences for $D\sim 1~{\rm m}$ would not be noticeable in this plot; intermediate shading indicates results for $\theta=90^\circ$. This illustrates the subdued blurring predicted in the far UV and X-rays, e.g. below the resolutions of instruments onboard {\it Swift}. Similarly, it does not preclude detection of $\gamma$-rays with {\it Fermi} LAT, $D\approx1.8$ m. It may be difficult to gain new observational insight at TeV and higher energies though, as $\Phi$ is close to the FoV of those telescopes. However, a value of $\Phi=34^{\circ}$ for $\theta=55^{\circ}$ at the GZK limit (not shown in Figure~\ref{figure_blurring_wide}) is comparable to an EeV hotspot already reported by \cite{Abbasi2014}, who averaged detections in $20^\circ$ bins for effectively $D\approx1.7$ m sensors of the AT.

It may be that Planck-scale-induced blurring is just at (or below) what is visible, so the usefulness of this revised $\Phi$ model could be in its generality. It should be possible to falsify at any observable wavelength where telescope resolution is better. An important region to try is evidently where $\Phi$ is predicted to ``turn over" close to the best resolution with {\it Fermi} LAT near 10 GeV. Further investigation of the available data there was pursued.

\begin{figure*}
\plotonewide{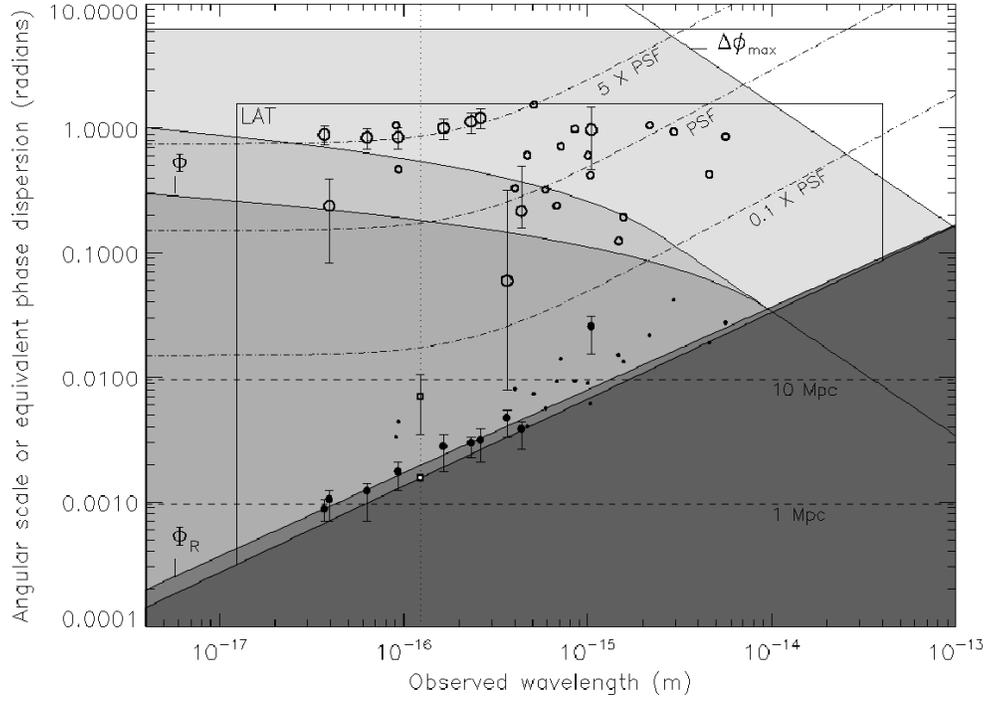}
\caption{Same as Figure~\ref{figure_blurring_wide} except for $z=0.5$ and confined to the region probed by {\it Fermi} LAT, darkly shaded below the 90\%-confidence resolution limit. Component $\Phi_{\rm R}$ is plotted with $R=1.9\times10^8$ and $\rho=0.70$, for $D=1.8~{\rm m}$. Open symbols indicate pointing angles, filled ones are limiting 90\%-confidence point-source identification resolutions. Larger symbols are for sources with known $z\geq0.5$. For these, angular errors are 1-$\sigma$ limits; resolutions are 68\% (lower) and 95\% (upper) confidence limits. For comparison, instrumental PSF limits are over-plotted as dot-dashed curves.}\label{figure_blurring_gamma}
\end{figure*}

\section{{\it Fermi} LAT, Data and Analysis}\label{observations}

The LAT is a pair-production imaging telescope, sensitive to $\gamma$-rays through resultant electron/positron tracks. Scintillometers covering the array are used to reject background events and a calorimeter allows recovery of incident $\gamma$-ray energy. The instrument details are described in \cite{Atwood2009}. Routine operation began in 2008; greater effective area and better resolution improves on the previous generation Energetic Gamma Ray Experiment Telescope (EGRET) on {\it Compton}.

Catalogs of positional accuracies of AGN and GRB source detections so far obtained with LAT have recently been released \citep{Ackermann2013a, Ackermann2013b}. These include 216 AGNs with a spectroscopically confirmed redshift, and 27 GRBs of which 10 have $z\geq0.5$. All the GRBs are plotted in Figure~\ref{figure_blurring_gamma}. Each is provided a peak energy of the detected $\gamma$-rays, which is here transformed to a single wavelength, scaling by $1/(2\pi c \hbar)$. Two distinct angular scales are provided in the catalogs. The first is effectively the apparent size of the source on the sky, which is related to the angle that the telescope must slew to bore-sight a source with the LAT once detected. The FoV of the LAT is $90^\circ$ wide with a PSF at 10 GeV over $10^\circ$ across \citep{Ackermann2013c}. The PSF has a uniform shape and so provides a well-defined centroid which can be confined to angles orders of magnitude smaller than its radius. This second resolution limit is actually a set of three numbers which are internal angular accuracies of centroid location on the sky: 68\%, 90\%, and 95\% confidence limits. Note that these positions are strictly from LAT, not by follow-up observations with another telescope, for example a UV or optical identification. The 90\% limit is adopted to represent the instrumental resolution, and will be used consistently hereafter. This allows the 68\% and 95\% limits to serve as boundaries on how well internal instrumental uncertainties are confined.

The PSF of LAT measured on orbit is indicated in Figure~\ref{figure_blurring_gamma} as dot-dashed curves, measured from low-$z$ AGNs and pulsars by \cite{Ackermann2013c} and given by their equation 1 for a 68\% containment radius. The energy dependence to the 90\%-confidence resolution limit, defined here as the most reliable centroiding accuracy possible, is well matched by a power law with slope: $\rho=\log{(5^\circ / 1.5\arcmin)}/\log{(60~{\rm GeV}/30~{\rm MeV})}=0.70$ and scaling $R=1.9\times10^8$; shown as a darkly-shaded region in Figure~\ref{figure_blurring_gamma}. Detailed scrutiny of the resolution limit came from searches for resolved 1-10 Mpc-sized haloes around $\gamma$-ray images of distant sources. The observed mean halo of AGNs described in \cite{Ando2010} and their resolution limit are plotted as two open squares (the halo shown with 1-$\sigma$ error bars). The apparent sizes of 1, 2, 5 and 10 Mpc scales are plotted for $z=0.5$. Low-redshift sources would be expected to be unresolved, and indeed a strict test can be performed on these: the average of all AGN 90\%-confidence limits for those with $z\leq0.5$. This is indicated by the white circle at 10 GeV, which agrees with both Ando et al. and Ackermann et al. results and so provides assurance that findings here are not in conflict. 

\begin{figure*}
\plotonewide{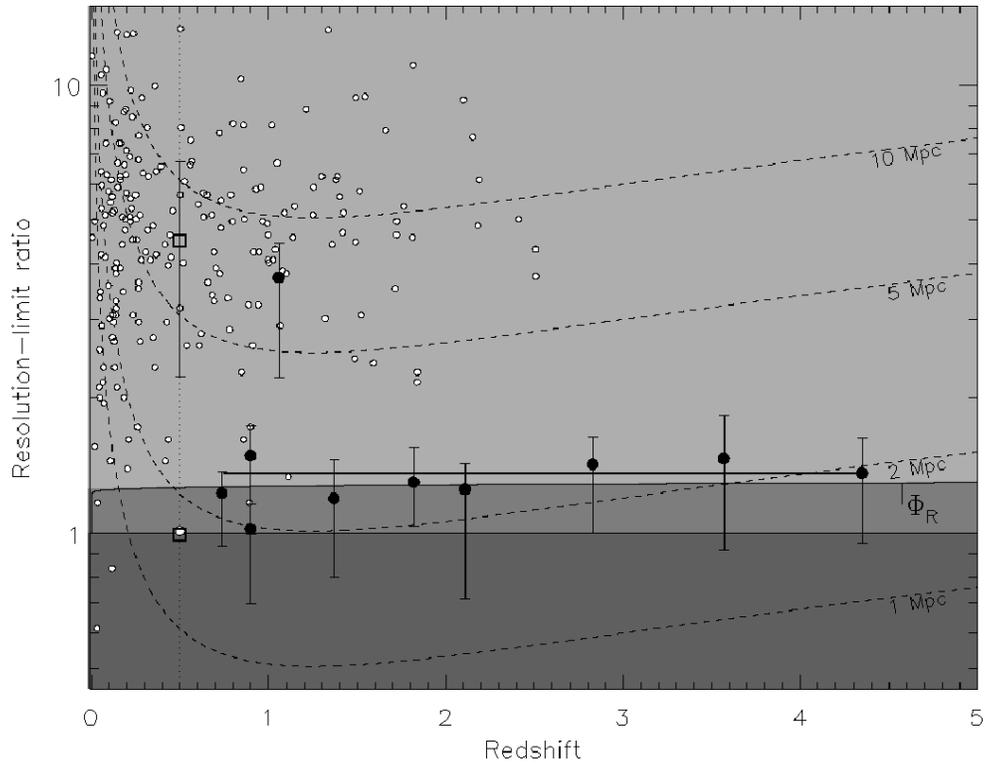}
\caption{{\it Fermi} LAT $z\geq0.5$ GRB position accuracies normalized by peak energy (filled dark symbols) plotted by redshift. The LAT AGN results are also shown (white). The dark shaded region is precluded by the 90\%-confidence resolution limit; lighter-shaded area that inflated as per $\Phi_{\rm R}$. Apparent angular scales are indicated by dashed curves, with a vertical dotted line at $z=0.5$.}
\label{figure_blurring_redshift}
\end{figure*}

Two correlations are of note in Figure~\ref{figure_blurring_gamma}. The first is that none of the GRB sources has a reported apparent angular size much smaller than $\Phi$, even for energies higher than 10 GeV, where the instrumental PSF is expected to improve on that. Only one exception shortward of 10 GeV is still within its uncertainty in this log-scaled plot. Note that most (6 of 10) $z\geq0.5$ sources follow a locus of the upper range of instrumental PSF width at their peak energies, none of which is found in the region formed by the limiting case of $\theta=2\pi$. This constrains $\alpha$, if $\Phi$ is correct. Even for $L$ reduced to the lower $L_{\rm C}$ this would be violated for a value of $\alpha\leq0.65$. The second interesting trend is that of centroiding accuracy: none of the $z\geq0.5$ GRBs has a 90\%-confidence accuracy better than the resolution limit of the telescope. On the contrary, one would expect that in the absence of Planck-scale effects true point-sources should all be uniformly along that line.

Possible enhancement of GRB positional uncertainty beyond what is expected just from the resolution limit is further illustrated in Figure~\ref{figure_blurring_redshift}. Here the ratios of positional accuracy to telescope resolution are plotted for the $z\geq0.5$ GRBs with respect to source redshift. All the AGN centroid 90\%-confidence accuracies, normalized to 10 GeV, are also plotted in the same way as small white dots; the minimum (and scatter) among those with $z\leq 0.5$ helps indicate how well telescope resolution is characterized with current best samples. For comparison, various apparent sizes (at 10 GeV) as a function of redshift are plotted as dashes. These are relatively constant in the range $0.5 \leq z \leq 4$. The GRB centroid accuracies do seem to match the $\Phi_{\rm R}$ prediction: sources with $z\geq0.5$ should not be located to better than $27\%$ over the telescope resolution, on average (for $a_0$, $\alpha$, $R$ and $\rho$ previously given). They are not, within their uncertainties, including an outlying source at $z=1.2$; their median is indicated by the thick horizontal line. Even greater blurring (up to the maximum) is not disallowed, i.e. an integration influenced by ``freak" photon events. But an alternate explanation that cannot be ruled out is the effect of partially resolved sources: a halo $\sim 5-10$~Mpc across for the outlier and plausibly a common 2 Mpc for the rest. Although just at the level of detection, previous LAT studies \cite[e.g.][]{Ackermann2013b} do not report a similar enhancement, so that may be of interest even if it does not reveal evidence of Planck-scale-induced blurring.

The recent Fermi detection of GRB 130427 is also interesting due to its unrivalled brightness across such a broad range of energies \citep{Ackermann2014}. In fact, at least one photon near 100 GeV arrived from this source. Unfortunately, it has a redshift of only $z=0.34$ which greatly reduces discrimination between the energy dependence of the LAT PSF, the action of Planck-scale blurring, and other competing scattering effects - so as to estimate how likely that is to happen. A higher-$z$ source may be telling, but below $z=0.5$ there is a dramatic upturn in length scales (apparent in Figure~\ref{figure_blurring_redshift}). A photon cannot be distinguished between blurring as per $\Phi_{\rm R}$ and a 2 Mpc halo for $z\leq0.5$.

Larger samples from {\it Fermi} will be increasingly restrictive. Newly detected $z>0.5$ GRBs with peak energy higher than 10 GeV are expected to appear too broad, as $\Phi$ is larger than the instrumental PSF there. The predicted enhancement in positional uncertainty of $\Phi_{\rm R}$ has almost no dependence on redshift (once that becomes significant) although intrinsic source size does, and so greater statistics could be used to disentangle these from systematic uncertainty in source position.

\section{Summary and Future Directions}\label{summary}

A treatment of Planck-scale-induced blurring of monochromatic image angular size has been presented. This can reconcile the best-available space-based measurements of among the highest-known redshift pointlike sources viewed in the optical and in $\gamma$-rays. Results are consistent with previous {\it HST} results: there is agreement on the weak effect for $3.9 \leq z \leq 6.3$ AGNs comparable to the limit reported before. The 10 currently known $0.5 \leq z \leq 4.3$ GRBs from {\it Fermi} follow a growth in apparent angular size with increasing energy (beyond what is expected from its PSF) consistent with the revised prescription, constrained to be $\alpha>0.65$ if $a_0=1$. There is potential for confusion due to $\gamma$-ray haloes. But there is also some reason to expect that the degeneracy can be broken with greater statistics, at least for GRBs. The generality of $\Phi$ may be helpful: based on this analysis, if one observed a $z=4$ {\it Fermi} GRB with the 6.5 m {\it James Webb Space Telescope} it would pass a limit of 10\% larger than diffraction shortward of 350 nm. Agreement across such a broad wavelength range may strengthen the confirmation by ruling out conspiring intrinsic source sizes being at blame, which could still hamper the case for AGNs.

At the high-energy extreme, wavefronts blurred beyond the horizon have been explicitly excluded, a necessary condition to explain the observability of pointlike $\gamma$-ray sources. That is a falsifiable assertion, as those blurred more than $\Phi$ in equation 7 must be lost to the background and so could still be ``observable" by their absence - if the expected luminosity of the source is well understood. One possibility is to investigate the GZK energy limit, as this is where this effect should be strongest. Current indications of anisotropy are not in conflict with the results here, but cosmic-ray energy distributions were not included, so more is beyond the scope of this paper. It may be a direction of future study, as could temporal variation via the statistical nature of photon counting, an issue which has been strictly avoided in this instance by considering only long-term averages. It is hoped that the current work will inspire further searches for Planck-scale-induced blurring with GRB follow-up observations. It is worth looking for: the import of a clear detection is nothing less than route towards a successful quantum gravity model.

\acknowledgements

I thank Richard Lieu for long ago noting to me the inconsistency of the strong limits to Planck-scale-induced blurring and detected $\gamma$-ray sources. Thoughtful comments from an anonymous referee helped improve the original manuscript.

\end{document}